\documentclass[preprint,12pt]{elsarticle}

\usepackage{amssymb}
\usepackage{amsmath}
\usepackage{multirow}

\journal{Elsevier}

\begin{document}

\begin{frontmatter}



\title{Anomalous Ionic Conductivity along the Coherent $\Sigma 3$ Grain Boundary in ThO$_2$}
\date{}
\author[inst1]{Miaomiao Jin\corref{cor1}}
\ead{mmjin@psu.edu}
\cortext[cor1]{Corresponding author}
\affiliation[inst1]{Department of Nuclear Engineering, The Pennsylvania State University, University Park, 16802 PA, USA}
\author[inst1]{Jilang Miao}
\author[inst2]{Marat Khafizov}%
\affiliation[inst2]{Department of Mechanical and Aerospace Engineering, The Ohio State University, 201 W 19th Ave, Columbus, OH 43210, USA}
\author[inst1]{Beihan Chen}
\author[inst3]{Yongfeng Zhang}%
\affiliation[inst3]{Department of Engineering Physics, University of Wisconsin-Madison, 1500 Engineering Dr, Madison, WI 53706, USA} 
\author[inst4]{David H. Hurley}%
\affiliation[inst4]{Idaho National Laboratory, 2525 Fremont Ave, Idaho Falls, ID 83402, USA}%

\begin{abstract}
Understanding oxygen diffusion along grain boundaries (GBs) is critical for controlling ionic conductivity in oxide ceramics. GBs are typically thought to enhance ionic transport due to structural disorder and increased free volume. In this study, we report an unexpected anomaly: the $\Sigma 3(111)$ GB in thorium dioxide (ThO$_2$), despite its compact and coherent structure, exhibits significantly higher oxygen ionic conductivity compared to the more open GB ($\Sigma 19$ as an example). Using atomistic simulations based on a machine learning interatomic potential, we revealed that the high conductivity in the $\Sigma 3$ GB arises from a collective diffusion mechanism involving highly correlated atomic motion reminiscent of a superionic state. In contrast, the $\Sigma 19$ GB follows conventional pipe diffusion, consistent with its more open structure. This comparison highlights that enhanced GB conductivity is not simply correlated with free volume, but can occur from specific structural motifs that enable collective transport. These findings provide new guidance for designing GB-engineered oxides with targeted ionic transport properties for energy applications.
\end{abstract}

\begin{keyword}

Oxygen diffusion \sep grain boundary \sep correlation \sep superionic

\end{keyword}

\end{frontmatter}


%\section{Introduction}
%\label{sec:sample1}

%%%%%%%%%%%%%%%%%%%%% Introduction %%%%%%%%%%%%%%%%%
 
Ionic conductivity in polycrystalline oxide materials is critical for applications in nuclear fuels, solid oxide fuel cells, and ion-conducting membranes \cite{williams2015atomistic,murphy2014pipe,vincent2009self,arima2010molecular,wang2024accelerating}. Grain boundaries (GBs) in these materials are key to enhancing or hindering ionic transport due to their structural and chemical heterogeneity \cite{wang2024accelerating,nakagawa2011grain,nakagawa2007yttrium,milan2022role}. While it is apparent that GB affects ionic conductivity, the specific relationship between GB structure and ionic conductivity in oxides remains an area of active research. For example, significantly enhanced oxygen diffusion in GB than in the bulk has been found in oxides such as UO$_2$ \cite{arima2010molecular,nishina2011molecular}, ThO$_2$ \cite{jin2024extended}, and Al$_2$O$_3$ \cite{heuer2013growth}. Nevertheless, other studies noted that the GB oxygen conductivities of doped ZrO$_2$ and CeO$_2$ \cite{guo2006electrical, de2008oxygen, gonzalez2012molecular} are much lower than the corresponding bulk values, due to severely reduced oxygen vacancies and impurity blocking the ionic transport at GB. Hence, these existing findings underscore the complex relationship between GB and ionic transport. 

The enhanced diffusion at GB is commonly explained by point defects hopping in the relatively open structure at GBs, particularly for high-angle GBs which are commonly perceived as sources/sinks for point defects \cite{vincent2009self,arima2010molecular,nishina2011molecular}. An alternative explanation was proposed in understanding oxygen diffusion in Al$_2$O$_3$, where such GB diffusion is thought to occur by migration of disconnections containing positively charged and negatively charged jogs for random high-angle grain boundaries, as an analogy to dislocation migration \cite{heuer2013growth,heuer2015disconnection}. Although discussions are ubiquitous on general GBs, this leaves a gap in understanding how the highly symmetrical, coherent, low-energy twin boundary contributes to ionic conductivity. In this case, the vacancy-mediated or GB-ledge-mediated diffusion would not work for oxygen transport along the twin GBs. Interestingly, our previous work indicates that $\Sigma 3(111)$ exhibits stronger oxygen diffusion than open-space GBs in ThO$_2$ \cite{jin2024extended}; it was suspected that in  $\Sigma 3(111)$ GB, there are other oxygen diffusion mechanisms involved, while the underlying reason was not elucidated.

%Feng et al. observed that $\Sigma 3$ twin GBs in CeO$_2$ strongly maintain oxygen stoichiometry \cite{feng2012atomic}, which suggests that $\Sigma$3 GB holds strong structural coherency with low vacancy concentration.

ThO$_2$, provides an ideal model system to investigate GB transport phenomena. Unlike other fluorite-structured oxides such as CeO2 or UO2, ThO2 lacks complications tied to presence of ground state $f$-electron, allowing more accurate modeling of defect energetics and diffusion mechanisms. For example,  this absence allows for reliable density functional theory (DFT) calculations with standard functionals, making the trained interatomic potential more robust for atomistic simulations of diffusion processes. Hence, in this work, we advance the previous work by turning to a recently developed machine learning potential for ThO$_2$ \cite{jiang2024machine}. Through extensive atomistic simulations, we have confirmed that the coherent and highly ordered $\Sigma 3$ (111) twin boundary displays markedly higher oxygen diffusivity than the higher-angle, structurally open GB. Detailed mechanistic analysis shows that this arises from a collective diffusion mode resembling a superionic state. Such correlated collective atom shuffling enable efficient oxygen transport with a lower activation barrier. Such understanding provides insights towards leveraging GB engineering to tune ionic conductivity in ThO$_2$ and similar materials.

%%%%%%%%%%%%%%%%%%%%% Methods %%%%%%%%%%%%%%%%%
Machine learning-enabled molecular dynamics (MD) simulations were performed using LAMMPS \cite{plimpton2007lammps} with a recently developed ML potential for ThO$_2$ \cite{jiang2024machine}, which has demonstrated excellent accuracy in reproducing defect energetics and diffusion properties. Two additional empirical interatomic potentials were also employed for comparison: the EAM potential by Cooper et al. \cite{cooper2014many} and the improved EAM potential by Zhou et al. \cite{zhou2025parameterizing}. A brief description of these empirical potentials is provided in the Supplementary Materials (SM). All simulation supercells were constructed to be charge-neutral. Simulations were performed over the temperature range of 2200 K to 2800 K, where sufficient atomic displacements occur on the MD timescale to ensure statistically significant diffusion measurements. For each temperature, eight independent simulations were conducted for statistical reliability. Within this temperature window, the oxygen sublattice in the bulk ThO$_2$ remains stable \cite{ghosh2016computational}, avoiding bulk superionic behavior. Two representative GB types were investigated: $\Sigma 3(111)/[\bar{1}10]$ (denoted as $\Sigma 3$) and $\Sigma 19(\bar{3}31)/[110]$ (denoted as $\Sigma 19$). The $\Sigma 3$ GB represents a highly compact and coherent interface, while the $\Sigma 19$ GB exhibits significant open-space character. The $\Sigma 3$ simulation cell contained 4,320 atoms with approximate dimensions of $2.70 \times 2.34 \times 9.60$ nm$^3$, while the $\Sigma 19$ cell contained 3,648 atoms with dimensions of $3.41 \times 1.56 \times 9.87$ nm$^3$; the cell volumes vary slightly with temperature due to thermal expansion. Oxygen diffusion coefficients ($D_{\mathrm{O}}$) were computed using both the mean squared displacement (MSD) method and the velocity autocorrelation function (VACF) method to capture both conventional and correlated diffusion behaviors. Spatially resolved diffusion coefficients were calculated as a function of distance from the GB center along the direction normal to the GB plane, following the methodology established in our previous work \cite{jin2024extended}. In addition, the van Hove correlation function, which captures both self-diffusion and interparticle correlations in a system, was computed to reveal dynamical correlations and possible collective motion of oxygen atoms. Finally, partial phonon density of states (PDOS) was extracted from the Fourier transform of velocity autocorrelation spectra to assess vibrational signatures associated with oxygen atoms.

%Additional simulation details and quantification of spatially dependent diffusivity are provided in the SM.  

\begin{figure}[!ht]
	\centering
	\includegraphics[width=0.95\textwidth]{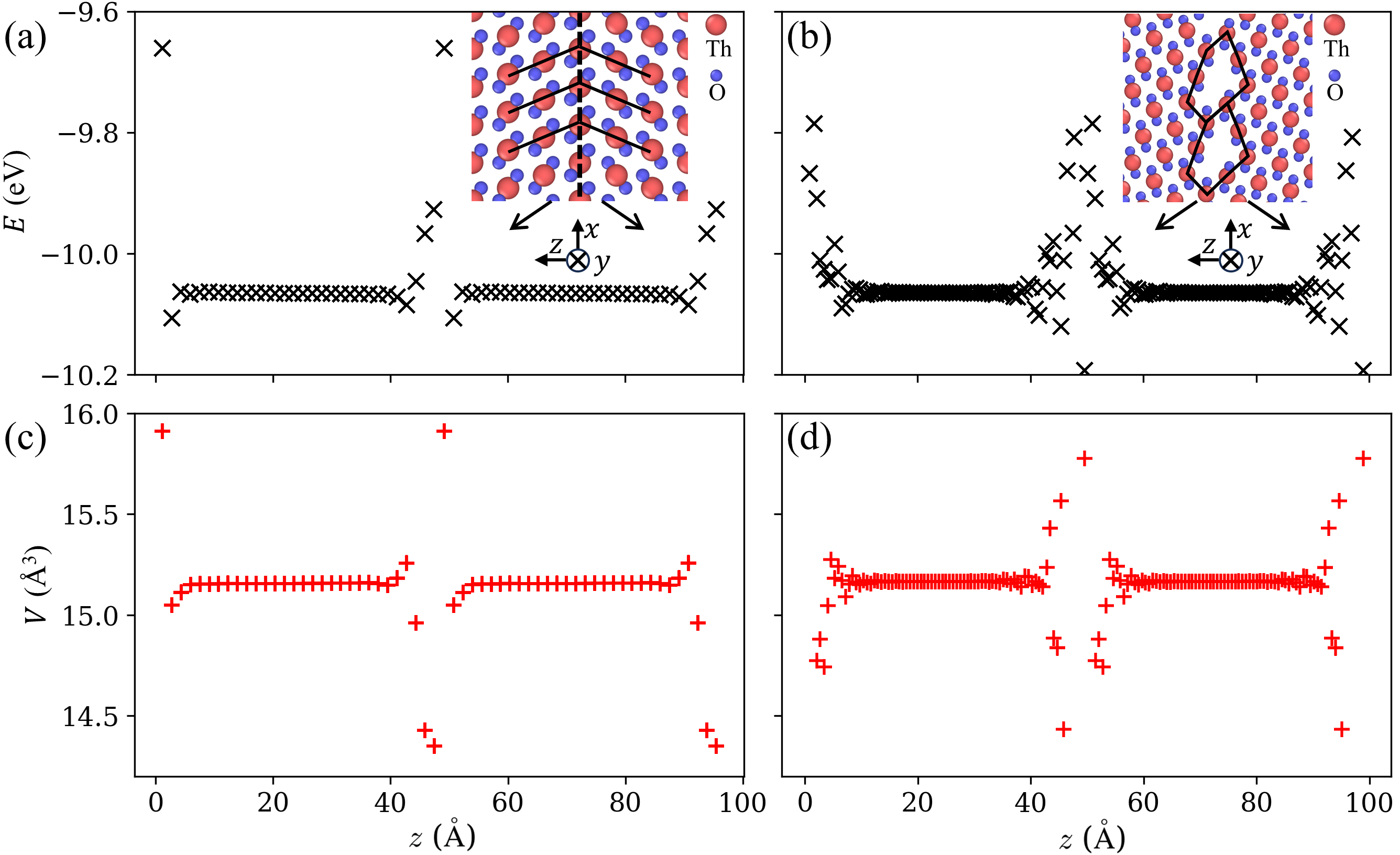}
	\caption{ Oxygen atomic energy and volume in $\Sigma 3(111)$ (a, c) and $\Sigma 19(331)$ (b, d) GB supercells. Horizontal axis indicates the $z$ coordinates of the atoms. Insets in (a) and (b) illustrate the GB structures.}
	\label{fig:GB}
\end{figure}

%%%%%%%%%%%%%%%%%%%%% Results %%%%%%%%%%%%%%%%%
The structural characteristics of the two GBs were first revealed by the oxygen atomic energies and volumes (based on Voronoi analysis), as shown in Figure~\ref{fig:GB}. The insets in Figure~\ref{fig:GB}a-b provide the views of the two GB structures after energy minimization, which are consistent with previous GB studies in UO$_2$ \cite{williams2015atomistic}. It is evident that the $\Sigma 3$ GB exhibits a compact and coherent structure, while the $\Sigma 19$ GB displays large open spaces along $y$-direction. In both GBs, oxygen atoms mostly exhibit higher energies compared to bulk oxygen atoms, though certain atoms, depending on local coordination, can adopt lower energy states. The $\Sigma 3$ GB exhibits a narrower energy distribution, though some oxygen atoms possess relatively high energy despite having larger volumes, which arises from the local packing at this coherent interface. In contrast, the $\Sigma 19$ GB shows greater variation in both energy and volume distributions, consistent with its structural disorder and open character. The atomic energy is generally negatively correlated with atomic volume, as shown in the SM. 
%Lastly, it can be inferred that the spatial width of the $\Sigma 3$ GB at ground state is much smaller than that of the $\Sigma 19$, suggesting more confined diffusion at the $\Sigma 3$ GB, as will be demonstrated in the following.

Complementing this local atomic analysis, we computed the grain boundary energy using the relation $(E_{\mathrm{GB}} - E_0) / (2A)$, where $E_{\mathrm{GB}}$ is the total potential energy of the system containing the GB, $E_0$ is the potential energy of a perfect lattice with the same number of atoms, and $A$ is the $x$-$y$ cross-sectional area of the simulation cell. The resulting GB energies are summarized in Table~\ref{tab:E_GB}, comparing values obtained from the ML potential and two empirical potentials (CRG \cite{cooper2014many} and Zhou \cite{zhou2025parameterizing}) under the same energy minimization settings.  The $\Sigma 3$ GB shows strong consistency across all three potentials, and also has a much lower energy than other GB types \cite{jin2024extended}. This is consistent with the conventional understanding across a wide range of materials, where coherent $\Sigma 3$ twin boundaries exhibit low interfacial energy and high structural stability.

%GB energy
\begin{table}[]
\centering
\caption{GB energy (J/m$^2$) calculated based on three interatomic potentials for ThO$_2$.}
\vspace{0.5em}
\label{tab:E_GB}
\begin{tabular}{|l|l|l|l|}
\hline
           & CRG \cite{cooper2014many} & Zhou \cite{zhou2025parameterizing} & MLIP \cite{jiang2024machine}\\ \hline
$\Sigma$3 GB &  0.88   &  1.00    &  1.03  \\ \hline
$\Sigma$19 GB &  1.69   &  3.10    &  1.59  \\ \hline
\end{tabular}
\end{table}

\begin{figure}[!ht]
	\centering
	\includegraphics[width=0.95\textwidth]{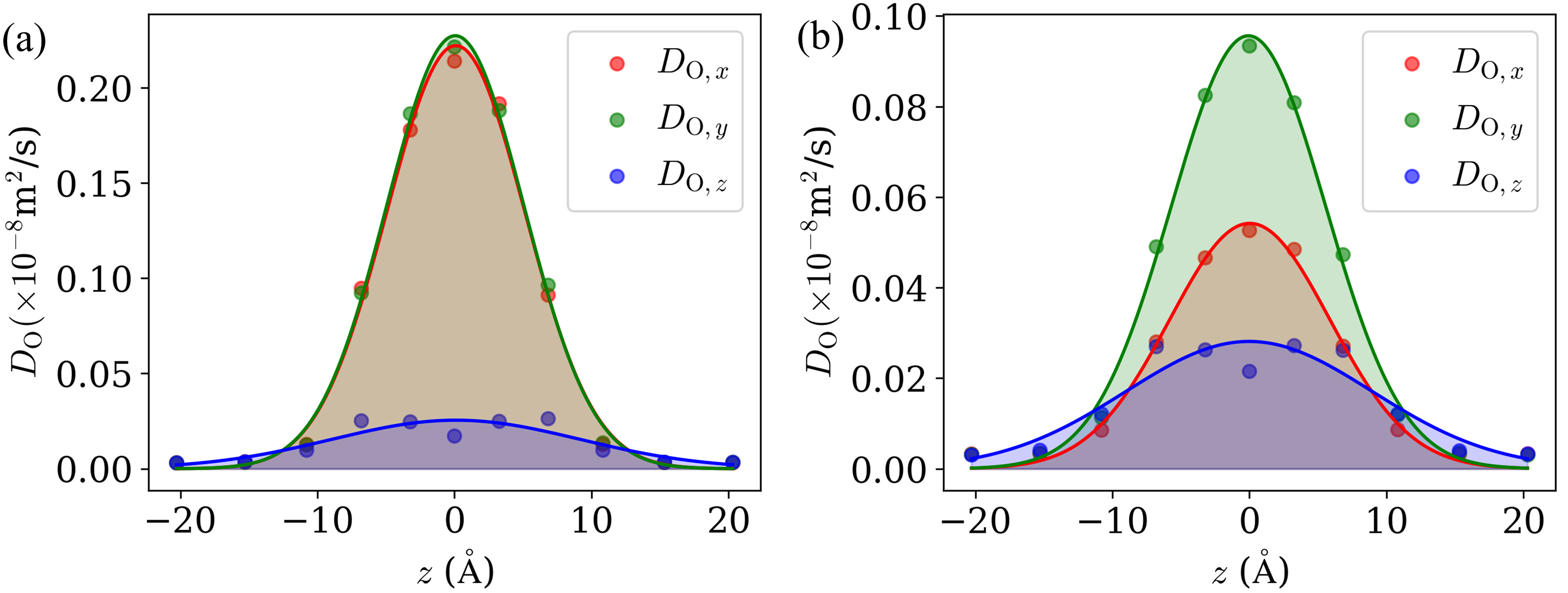}
	\caption{Spatially resolved oxygen diffusion coefficients at 2500 K for $\Sigma$3 (a) and $\Sigma$19 (b) GBs. GB center is position at 0.0 \AA.}
	\label{fig:GB_D}
\end{figure}

Next, we analyzed the spatially resolved oxygen diffusion behavior across the two GBs. Traditional models of GB diffusion suggest that diffusivity should be enhanced in boundaries that exhibit greater structural disorder and free volume, as these characteristics are expected to lower migration barriers and create interconnected diffusion pathways \cite{turnbull1954effect,balluffi1982grain,sorensen2000diffusion}. The inset of Figure~\ref{fig:GB}(b) shows that the $\Sigma 19$ GB contains large open channels along the $y$-direction, suggesting that this boundary may support pipe diffusion. In contrast, the $\Sigma 3$ GB is a compact boundary where one might expect diffusion to be suppressed while the results  suggest the opposite. Figure~\ref{fig:GB_D}  displays the spatially resolved oxygen diffusion coefficient $D_{\mathrm{O},\alpha}$ as a function of distance from the GB center at 2500 K, where $\alpha$ represents diffusion along the $x$, $y$, and $z$ directions. Since the simulated GBs are symmetric tilt boundaries, the diffusivity profiles are as expected symmetric with respect to the GB plane located $z=0$ {\AA}. In both GB types, oxygen diffusion is substantially enhanced in the immediate GB region compared to the bulk, consistent with prior studies of GB diffusion in fluorite oxides \cite{jin2024extended,williams2015atomistic,vincent2009self,arima2010molecular}. In the case of the $\Sigma 3$ GB, oxygen diffusion is isotropic within the GB plane, as indicated by the nearly overlapping $D_{\mathrm{O},x}$ and $D_{\mathrm{O},y}$ profiles. By comparison, diffusion perpendicular to the GB plane ($D_{\mathrm{O},z}$) is significantly lower, which reflects the confined nature of the GB.  Notably, it can be concluded here that the $\Sigma 3$ GB acts as a two-dimensional fast-diffusion channel within the three-dimensional grains. In contrast, the $\Sigma 19$ GB exhibits an anisotropic diffusion behavior. The enhancement in $D_{\mathrm{O}}$ is greatest along the $y$ direction, which aligns with the open structural channels. Overall, $D_{\mathrm{O},y} > D_{\mathrm{O},x} > D_{\mathrm{O},z}$ is observed. Such anisotropic behavior is commonly expected in GBs with structural channels for pipe diffusion \cite{turnbull1954effect,peterson1983grain}. 

Focusing on the GB regions, Figure~\ref{fig:GB_D} presents the temperature dependence of oxygen diffusivity, with $D_{\mathrm{O}}$ calculated from 2200 K to 2800 K. In all cases, oxygen diffusion exhibits Arrhenius behavior. We then extracted the activation energies. For the $\Sigma 3$ GB, oxygen diffusion within the GB plane ($D_\mathrm{O,x}$ and $D_\mathrm{O,y}$) displays approximately 0.8 eV (Table~\ref{tab:my-table}), consistent with the values derived from the CRG empirical potential (0.9 eV) \cite{jin2024extended}. Meanwhile, diffusion perpendicular to the GB plane ($D_\mathrm{O,z}$) shows a higher barrier of 1.3 eV, indicating the geometric confinement from the GB structure.  For the $\Sigma 19$ GB, these temperature-dependent values also reflect the structural channels along the $y$ direction. Although the open space in $\Sigma 19$ suggest an effective high-concentration of vacancy, the extracted activation energies are substantially higher than the bulk oxygen vacancy migration energies reported previously (2.4 eV versus 0.78 eV \cite{ghosh2016computational,colbourn1983calculated} and 0.64 eV \cite{he2022dislocation}). This is explained by the correlation dynamics. As shown in the SM, oxygen diffusivities calculated from the VACF are notably lower than those from MSD analysis, which indicates temporal correlations in the oxygen motion. These correlations likely arise from the stable cation lattice, which mediates oxygen migration. Hence, the $\Sigma 19$ GB supports a partially correlated vacancy hopping diffusion mechanism. The correlation elevates the apparent activation energy relative to simple point defect migration. As an argument, bulk diffusion (evaluated from regions 2 nm away from the GB plane) yields an activation energy of approximately 4.0 eV for both GB cases. However, this value is higher than that predicted by the indirect method: $0.5 E_{f,\mathrm{O-FP}} + E^m_{\mathrm{V_O}}$, which considers vacancy formation and migration \cite{murch1987oxygen,arima2010molecular}. Using $E_{f,\mathrm{O-FP}} = 4.36$ eV derived from this ML potential (computed by displacing an oxygen atom to a far-apart octahedral site) and $E^m_{\mathrm{V_O}} = 0.78$ eV \cite{ghosh2016computational}, \cite{colbourn1983calculated}, this approach predicts an effective activation energy of 2.96 eV. The discrepancy likely stems from the correlated oxygen motion in the dynamic simulations, which is not captured in the idealized Frenkel-pair model for diffusion analysis.

\begin{figure}[!ht]
	\centering
	\includegraphics[width=0.95\textwidth]{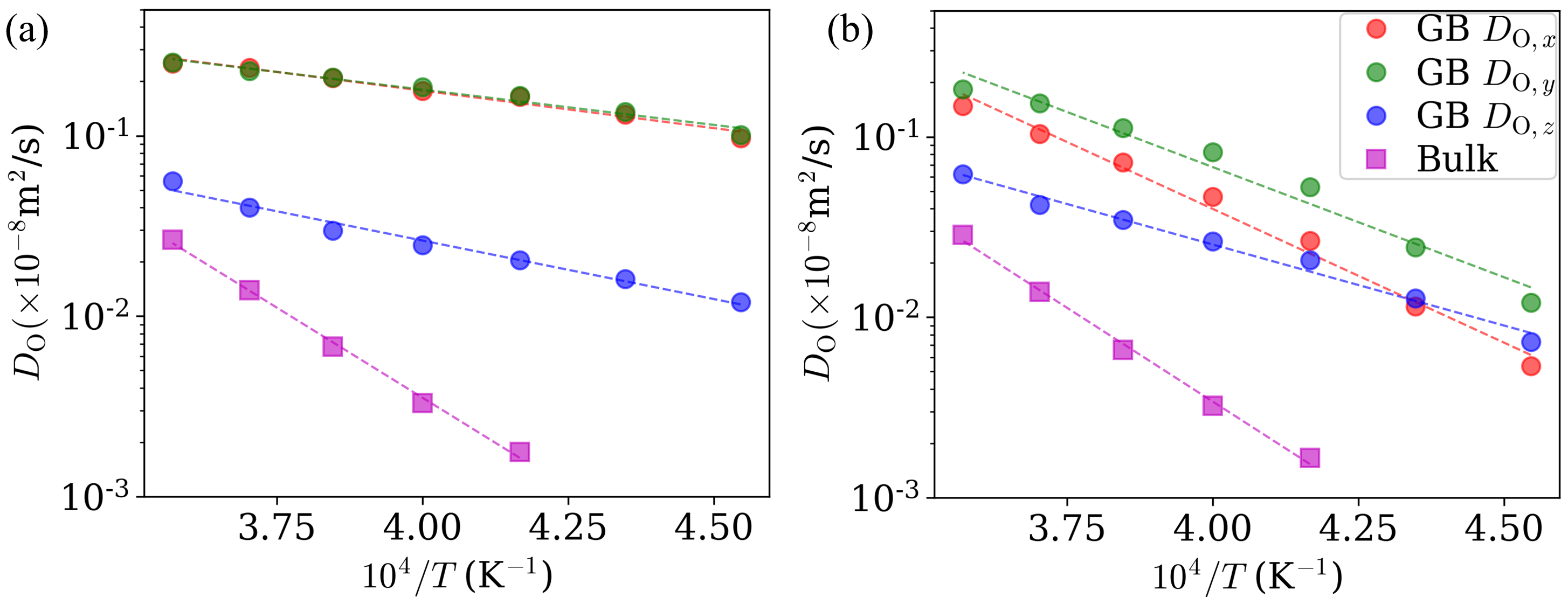}
	\caption{Oxygen diffusion coefficients in the GB and bulk regions as a function of inverse temperature, shown along the $x$, $y$, and $z$ directions for $\Sigma$3 GB (a) and $\Sigma$19 GB (b). The bulk values (extracted 2 nm away from the GB plane) are averaged across the three directions due to their highly comparable values.}
	\label{fig:GB_D_T}
\end{figure}
\begin{table}[]
\centering
\caption{Oxygen diffusion activation energy (eV) in the GB and bulk regions.}
\vspace{0.5em}
\label{tab:my-table}
\begin{tabular}{|l|l|l|l|l|}
\hline
            & $x$    & $y$    & $z$    & Bulk \\ \hline
$\Sigma 3$  & 0.8 & 0.8 & 1.3 & 4.0 \\ \hline
$\Sigma 19$ & 3.0 & 2.4 & 1.8 & 4.1\\ \hline
\end{tabular}
\end{table}
%$\Sigma 3$  & 0.82 & 0.78 & 1.29 & 3.97 \\ \hline
%$\Sigma 19$ & 2.95 & 2.43 & 1.78 & 4.13 \\ \hline

Note the comparison of the absolute magnitudes of diffusion coefficients between the two GBs: the $\Sigma 3$ GB exhibits a much higher overall diffusivity than that of the $\Sigma 19$ GB (especially on the lower temperature end), even though the latter features obvious open channels (this trend was also reported in UO$_2$ GBs \cite{williams2015atomistic}). The observed high oxygen diffusivity and low activation energy in $\Sigma 3$ GB indicate that oxygen transport occurs via alternative mechanisms, which are fundamentally different from that in $\Sigma 19$, as the perspective of free volume (enhanced vacancy-atom exchange) is insufficient to explain this behavior. In $\Sigma 3(111)$ GB, there are no inherent vacancies and interstitials in the ground state.  As will be further discussed below, the $\Sigma 3$ GB facilitates a superionic state in which oxygen ions undergo highly correlated motion confined to the GB plane. Such a coherent interface effectively acts as a ``superionic highway”, enabling significantly enhanced interfacial ionic conductivity. This collective transport arising from the superionic state is even more efficient than the pipe diffusion processes operating in $\Sigma 19$. 
 
\begin{figure}[!ht]
	\centering
	\includegraphics[width=0.95\textwidth]{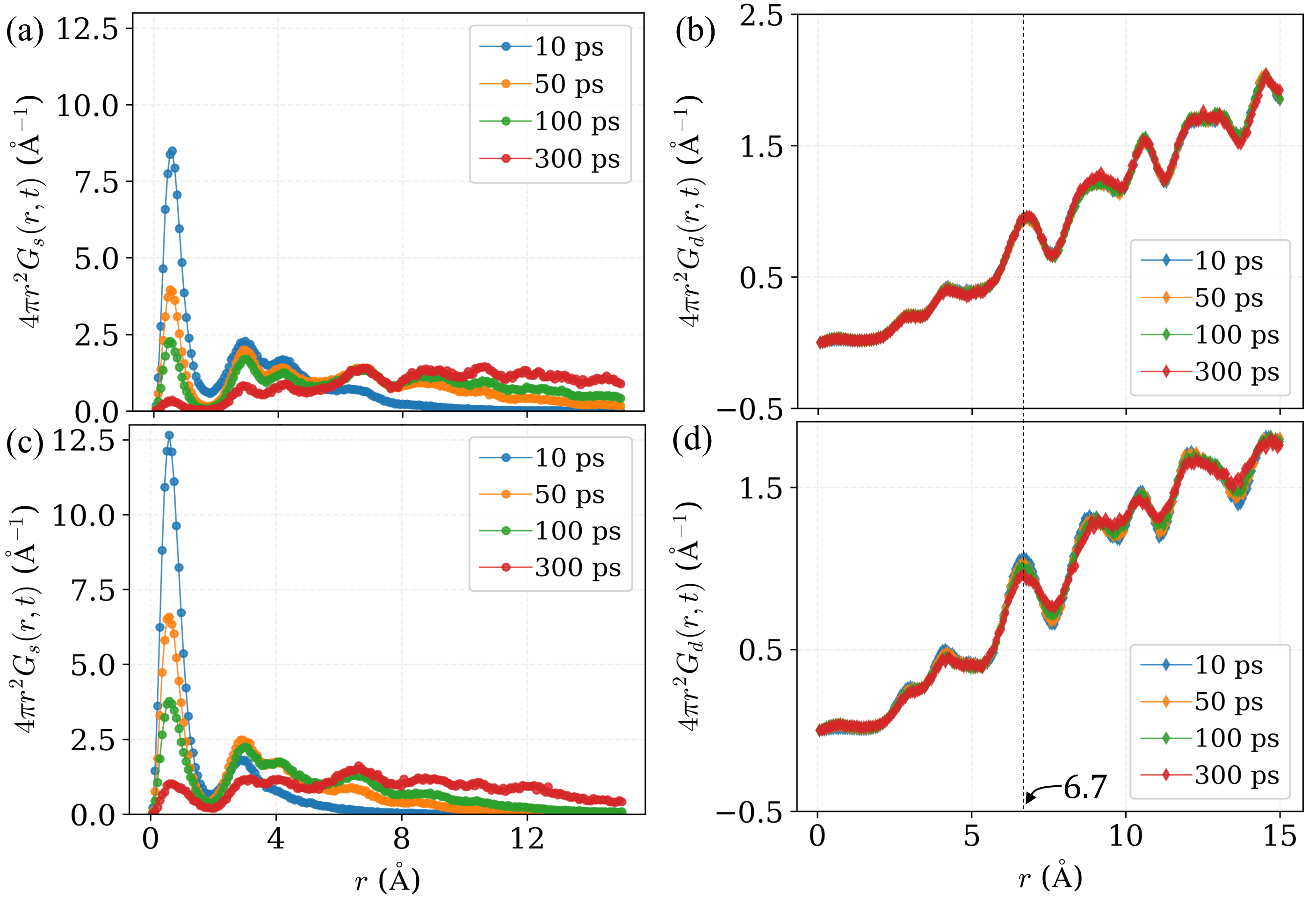}
	\caption{(a-b) show the self-part and distinct-part of the van Hove function of the $\Sigma$3 GB at 2500 K at 10, 50, 100, and 300 ps. (c-d) show the self-part and distinct-part of the van Hove function of the $\Sigma$19 GB at 2500 K at 10, 50, 100, and 300 ps.}
	\label{fig:van_Hove}
\end{figure}

To reveal the microscopic diffusion mechanisms, we first computed the van Hove correlation function at 2500 K, as shown in Figure~\ref{fig:van_Hove}, which provides insights into both single-particle and cooperative dynamics. The self-part of the van Hove function, $G_s(r,t)$, describes the probability that an oxygen atom initially at position $\mathbf{r}_0$ moves a distance $r$ after a time $t$, which reflects the rate of diffusion processes. In the case of the $\Sigma 3$ GB, $G_s(r,t)$ exhibits a rapid decay of the initial peak and a propagation of diffusion peaks to longer distances as time increases (Figure~\ref{fig:van_Hove}a), while for $\Sigma 19$ GB, $G_s(r,t)$ decays more slowly with slower dynamics (Figure~\ref{fig:van_Hove}c). Figure~\ref{fig:van_Hove}b and \ref{fig:van_Hove}d show the distinct-part of the van Hove function, $G_d(r,t)$, which captures spatial correlations between diffusing oxygen atoms. For the $\Sigma 3$ GB, $G_d(r,t)$ exhibits persistent peaks that remain stable over hundreds of picoseconds. These long-lived correlations suggest that groups of oxygen atoms are moving in a cooperative fashion, maintaining spatial relationships as they diffuse. As comparison, the distinct-part $G_d(r,t)$ for $\Sigma 19$ GB, shows peak broadening over time, due to gradual decorrelation of atomic positions during diffusion.

 \begin{figure}[!ht]
	\centering
	\includegraphics[width=0.95\textwidth]{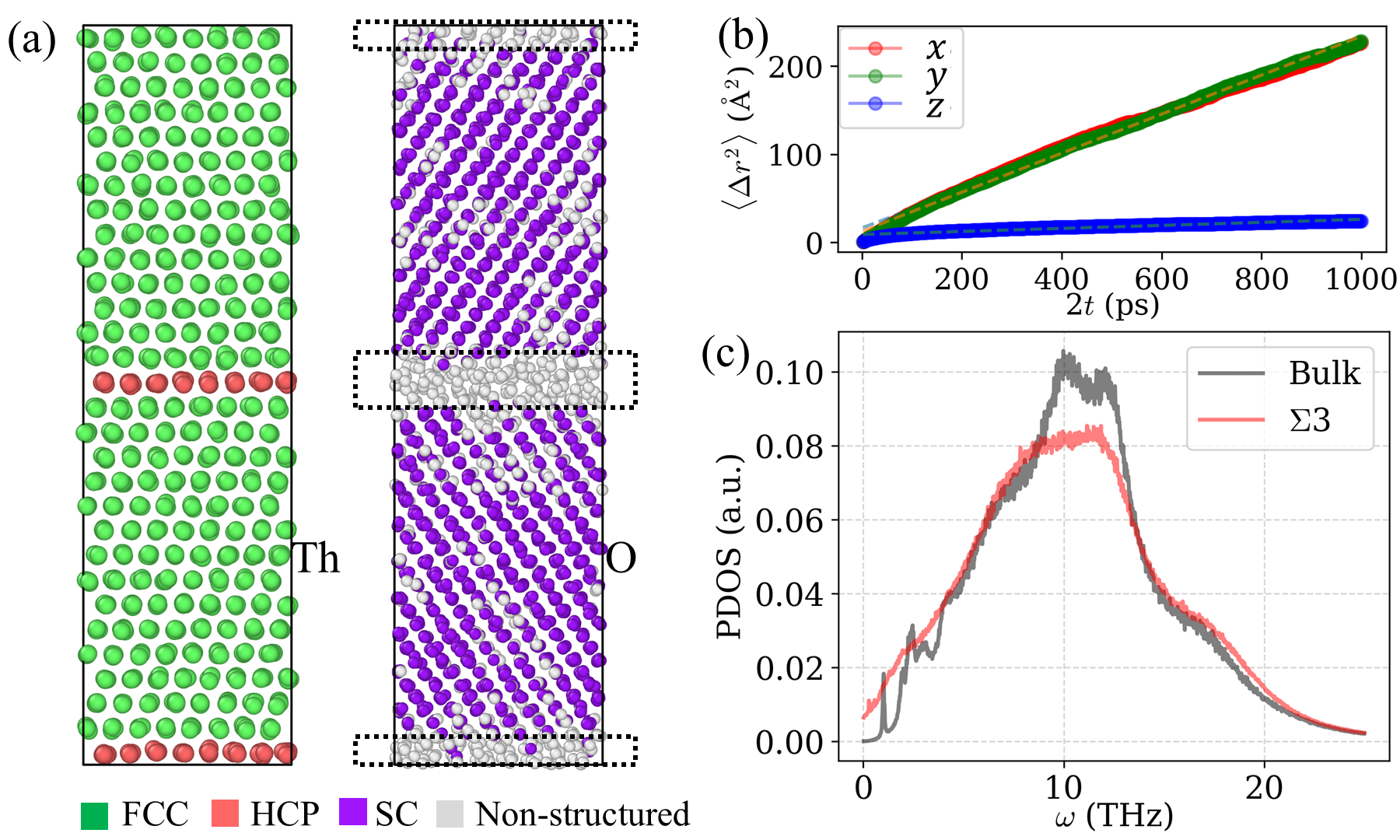}
	\caption{(a) Atomic configuration of the $\Sigma 3$ GB supercell at 2500 K. The left panel shows the Th sublattice; the right panel shows the O sublattice. Oxygen atoms are colored by local structural type: FCC (face-centered cubic), HCP (hexagonal close-packed), and SC (simple cubic). (b) Oxygen mean squared displacement along $x$, $y$, and $z$ directions within the $\Sigma 3$ GB region (highlighted by the dashed box in (a)) at 2500 K. (c) Partial phonon density of states (oxygen) comparing the $\Sigma 3$ GB and bulk regions at 2500 K. }
	\label{fig:GB_WHM}
\end{figure}

Further evidence for the superionic  state in the $\Sigma 3$ GB comes from direct visualization of atomic configurations and the analysis of vibrational dynamics. Figure~\ref{fig:GB_WHM}(a) provides a snapshot of the atomic structure after equilibration for 500 ps at 2500 K. The Th sublattice remains well-ordered as face-centered-cubic structure (interface shown as hexagonal close-packed atoms), which means that the high oxygen diffusivity is not due to disordering of GB structure defined by the cation lattice. Instead, the oxygen atoms in the $\Sigma 3$ GB region exhibit strong disorder, losing long-range order. Superionic transitions are well known in fluorite compounds at high temperatures, which features a sharp increase in ionic conductivity \cite{zhu2008theoretical,hutchings1987high,hutchings1984investigation}. In UO$_2$, the superionic transition is reported to occur around 2500--2670 K ($\sim$0.85 $T_m$) \cite{hutchings1987high,hiernaut1993premelting}. Given the higher melting point of ThO$_2$, the corresponding transition temperature is expected to be above 2500 K. Indeed, simulation studies using the CRG potential suggest a superionic transition near 2900 K for ThO$_2$ \cite{cooper2014thermophysical}, and the present simulations confirm that the bulk oxygen sublattice remains ordered at 2500 K. 

However, the presence of the $\Sigma 3$ GB effectively lowers the local superionic transition temperature, enabling a superionic state at the GB while the bulk remains crystalline state. From the MSD shown in Figure~\ref{fig:GB_WHM}b, the MSDs for $x$ and $y$ directions (within the GB plane) grow rapidly with time, while diffusion along the $z$ direction (perpendicular to the GB plane), remains significantly lower. The strong in-plane isotropy, combined with near-liquid diffusivity within the GB plane, indicates a quasi-two-dimensional superionic layer confined to the GB. Additionally, based on the distinct-part of the van Hove function in Figure~\ref{fig:van_Hove}b, the peaks in $G_d(r,t)$ at the $\Sigma 3$ GB exhibit no shifting. It suggests that oxygen ions are not diffusing completely liquid-like, but instead the diffusion is templated by the underlying cation lattice. Such structurally anchored superionic state is distinct from fully disordered regimes where $G_d(r,t)$ peaks typically shift as atomic correlations are lost. 

Finally, the PDOS for GB oxygen atoms offers a complementary understanding from the vibrational properties. As shown in Figure~\ref{fig:GB_WHM}c, comparing to the bulk lattice, in the $\Sigma 3$ GB region, the partial PDOS exhibits several features: i) there is pronounced enhancement of low-frequency acoustic modes, which promotes collective migration of oxygen in the GB region. ii) the high-frequency optical modes associated with oxygen vibrations are significantly reduced in intensity due to disordering of oxygen sublattice. iii) the optical region of the PDOS exhibits obvious peak smearing, reflecting strong anharmonic dynamics. Those observations thus further reflect that the $\Sigma 3$ GB supports a quasi-superionic state. 

In summary, the contrast between the two GB types suggests that while the enhanced diffusion in $\Sigma 19$ arises from local structural disorder and pipe diffusion, the $\Sigma 3$ GB supports a different transport mechanism based on dynamic correlations and collective ion migration. Specifically, the $\Sigma 3$ GB in ThO$_2$ hosts an quasi-superionic state confined to the GB plane, where oxygen ions migrate collectively with high mobility, while the Th cation sublattice remains crystalline. This behavior represents a key difference from conventional vacancy-mediated diffusion and highlights the unique role of low-$\Sigma$ coherent GBs in facilitating efficient ion transport. Importantly, the combination of high planar diffusivity, structural confinement, and vibrational softening not only advance our understanding of GB-enhanced ion transport in ThO$_2$ but also suggest opportunities for exploiting low-$\Sigma$ GBs to engineer enhanced ionic transport in functional ceramics.

\section*{Acknowledgments}
J. Miao is funded by The Pennsylvania State University. The rest co-authors are supported by the Center for Thermal Energy Transport under Irradiation, an Energy Frontier Research Center funded by the U.S. Department of Energy, Office of Science, United States, Office of Basic Energy Sciences. 
 
\section*{Data availability}
Data will be made available on request.

 \bibliographystyle{elsarticle-num} 
 \bibliography{cas-refs}

\begin{thebibliography}{10}
\expandafter\ifx\csname url\endcsname\relax
  \def\url#1{\texttt{#1}}\fi
\expandafter\ifx\csname urlprefix\endcsname\relax\def\urlprefix{URL }\fi
\expandafter\ifx\csname href\endcsname\relax
  \def\href#1#2{#2} \def\path#1{#1}\fi

\bibitem{williams2015atomistic}
N.~R. Williams, M.~Molinari, S.~C. Parker, M.~T. Storr, Atomistic investigation of the structure and transport properties of tilt grain boundaries of {UO2}, Journal of Nuclear Materials 458 (2015) 45--55.

\bibitem{murphy2014pipe}
S.~T. Murphy, E.~E. Jay, R.~W. Grimes, Pipe diffusion at dislocations in {UO2}, Journal of Nuclear Materials 447~(1-3) (2014) 143--149.

\bibitem{vincent2009self}
E.~Vincent-Aublant, J.~Delaye, L.~Van~Brutzel, Self-diffusion near symmetrical tilt grain boundaries in {UO2} matrix: A molecular dynamics simulation study, Journal of Nuclear Materials 392~(1) (2009) 114--120.

\bibitem{arima2010molecular}
T.~Arima, K.~Yoshida, K.~Idemitsu, Y.~Inagaki, I.~Sato, Molecular dynamics analysis of diffusion of uranium and oxygen ions in uranium dioxide, in: IOP Conference Series: Materials Science and Engineering, Vol.~9, IOP Publishing, 2010, p. 012003.

\bibitem{wang2024accelerating}
Y.~Wang, Z.~Chen, K.~Jiang, Z.~Shen, S.~Passerini, M.~Chen, Accelerating the development of llzo in solid-state batteries toward commercialization: A comprehensive review, Small (2024) 2402035.

\bibitem{nakagawa2011grain}
T.~Nakagawa, H.~Nishimura, I.~Sakaguchi, N.~Shibata, K.~Matsunaga, T.~Yamamoto, Y.~Ikuhara, Grain boundary character dependence of oxygen grain boundary diffusion in $\alpha$-al2o3 bicrystals, Scripta Materialia 65~(6) (2011) 544--547.

\bibitem{nakagawa2007yttrium}
T.~Nakagawa, I.~Sakaguchi, N.~Shibata, K.~Matsunaga, T.~Mizoguchi, T.~Yamamoto, H.~Haneda, Y.~Ikuhara, Yttrium doping effect on oxygen grain boundary diffusion in $\alpha$-{Al2O3}, Acta Materialia 55~(19) (2007) 6627--6633.

\bibitem{milan2022role}
E.~Milan, M.~Pasta, The role of grain boundaries in solid-state li-metal batteries, Materials Futures 2~(1) (2022) 013501.

\bibitem{nishina2011molecular}
M.~Nishina, K.~Yoshida, T.~Arima, Y.~Inagaki, K.~Idemitsu, I.~Sato, Molecular dynamics study on grain boundary diffusion of actinides and oxygen in oxide fuels, Progress in Nuclear Science and Technology 2 (2011) 95--100.

\bibitem{jin2024extended}
M.~Jin, J.~Miao, B.~Chen, M.~Khafizov, Y.~Zhang, D.~H. Hurley, Extended defects-enhanced oxygen diffusion in {ThO2}, Computational Materials Science 235 (2024) 112842.

\bibitem{heuer2013growth}
A.~Heuer, T.~Nakagawa, M.~Azar, D.~Hovis, J.~Smialek, B.~Gleeson, N.~Hine, H.~Guhl, H.-S. Lee, P.~Tangney, et~al., On the growth of {Al2O3} scales, Acta Materialia 61~(18) (2013) 6670--6683.

\bibitem{guo2006electrical}
X.~Guo, R.~Waser, Electrical properties of the grain boundaries of oxygen ion conductors: acceptor-doped zirconia and ceria, Progress in Materials Science 51~(2) (2006) 151--210.

\bibitem{de2008oxygen}
R.~A. De~Souza, M.~J. Pietrowski, U.~Anselmi-Tamburini, S.~Kim, Z.~A. Munir, M.~Martin, Oxygen diffusion in nanocrystalline yttria-stabilized zirconia: the effect of grain boundaries, Physical Chemistry Chemical Physics 10~(15) (2008) 2067--2072.

\bibitem{gonzalez2012molecular}
R.~L. Gonz{\'a}lez-Romero, J.~J. Mel{\'e}ndez, D.~G{\'o}mez-Garc{\'\i}a, F.~L. Cumbrera, A.~Dom{\'\i}nguez-Rodr{\'\i}guez, A molecular dynamics study of grain boundaries in {YSZ}: Structure, energetics and diffusion of oxygen, Solid State Ionics 219 (2012) 1--10.

\bibitem{heuer2015disconnection}
A.~Heuer, M.~Z. Azar, A disconnection mechanism of enhanced grain boundary diffusion in {Al2O3}, Scripta Materialia 102 (2015) 15--18.

\bibitem{jiang2024machine}
C.~Jiang, C.~A. Marianetti, M.~Khafizov, D.~H. Hurley, Machine learning potential assisted exploration of complex defect potential energy surfaces, npj Computational Materials 10~(1) (2024) 21.

\bibitem{plimpton2007lammps}
S.~Plimpton, P.~Crozier, A.~Thompson, {LAMMPS}-large-scale atomic/molecular massively parallel simulator, Journal of Computational Physics 18 (2007).

\bibitem{cooper2014many}
M.~Cooper, M.~Rushton, R.~Grimes, A many-body potential approach to modelling the thermomechanical properties of actinide oxides, Journal of Physics: Condensed Matter 26~(10) (2014) 105401.

\bibitem{zhou2025parameterizing}
S.~Zhou, C.~Jiang, E.~Xiao, S.~Bandi, M.~Cooper, M.~Jin, D.~H. Hurley, M.~Khafizov, C.~A. Marianetti, Parameterizing empirical interatomic potentials for predicting thermophysical properties via an irreducible derivative approach: the case of {ThO2} and {UO2}, Journal of Physics: Condensed Matter (2025).

\bibitem{ghosh2016computational}
P.~Ghosh, A.~Arya, G.~Dey, N.~Kuganathan, R.~Grimes, A computational study on the superionic behaviour of {ThO2}, Physical Chemistry Chemical Physics 18~(46) (2016) 31494--31504.

\bibitem{turnbull1954effect}
D.~Turnbull, R.~Hoffman, The effect of relative crystal and boundary orientations on grain boundary diffusion rates, Acta Metallurgica 2~(3) (1954) 419--426.

\bibitem{balluffi1982grain}
R.~W. Balluffi, Grain boundary diffusion mechanisms in metals (1982).

\bibitem{sorensen2000diffusion}
M.~R. S{\o}rensen, Y.~Mishin, A.~F. Voter, Diffusion mechanisms in cu grain boundaries, Physical Review B 62~(6) (2000) 3658.

\bibitem{peterson1983grain}
N.~Peterson, Grain-boundary diffusion in metals, International metals reviews 28~(1) (1983) 65--91.

\bibitem{colbourn1983calculated}
E.~Colbourn, W.~Mackrodt, The calculated defect structure of thoria, Journal of Nuclear Materials 118~(1) (1983) 50--59.

\bibitem{he2022dislocation}
L.~He, T.~Yao, K.~Bawane, M.~Jin, C.~Jiang, X.~Liu, W.-Y. Chen, J.~M. Mann, D.~H. Hurley, J.~Gan, et~al., Dislocation loop evolution in {Kr}-irradiated {ThO2}, Journal of the American Ceramic Society 105~(8) (2022) 5419--5435.

\bibitem{murch1987oxygen}
G.~Murch, C.~R.~A. Catlow, Oxygen diffusion in {UO2}, {ThO2} and {PuO2}. a review, Journal of the Chemical Society, Faraday Transactions 2: Molecular and Chemical Physics 83~(7) (1987) 1157--1169.

\bibitem{zhu2008theoretical}
B.~Zhu, S.~Li, B.-E. Mellander, Theoretical approach on ceria-based two-phase electrolytes for low temperature (300--600° c) solid oxide fuel cells, Electrochemistry Communications 10~(2) (2008) 302--305.

\bibitem{hutchings1987high}
M.~T. Hutchings, High-temperature studies of {UO2} and {ThO2} using neutron scattering techniques, Journal of the Chemical Society, Faraday Transactions 2: Molecular and Chemical Physics 83~(7) (1987) 1083--1103.

\bibitem{hutchings1984investigation}
M.~Hutchings, K.~Clausen, M.~Dickens, W.~Hayes, J.~Kjems, P.~Schnabel, C.~Smith, Investigation of thermally induced anion disorder in fluorites using neutron scattering techniques, Journal of Physics C: Solid State Physics 17~(22) (1984) 3903.

\bibitem{hiernaut1993premelting}
J.~Hiernaut, G.~Hyland, C.~Ronchi, Premelting transition in uranium dioxide, International journal of thermophysics 14 (1993) 259--283.

\bibitem{cooper2014thermophysical}
M.~W. Cooper, S.~T. Murphy, P.~C. Fossati, M.~J. Rushton, R.~W. Grimes, Thermophysical and anion diffusion properties of {(Ux,Th1-x)O2}, Proceedings of the Royal Society A: Mathematical, Physical and Engineering Sciences 470~(2171) (2014) 20140427.

\end{thebibliography}





% \end{thebibliography}
\end{document}